\documentclass[prl,showpacs,superscriptaddress,twocolumn]{revtex4-1}

\usepackage[makeroom]{cancel}

\usepackage{amsmath}    
\usepackage{amssymb}
\usepackage{graphicx}   
\usepackage{verbatim}   
\usepackage{color}      
\usepackage{hyperref}   
\usepackage[normalem]{ulem}
\usepackage{natbib}
\usepackage{enumitem}
\usepackage{empheq}

\hypersetup{colorlinks,linkcolor=blue,urlcolor=blue,citecolor=blue}

\definecolor{darkgreen}{rgb}{0,0.5,0}
\definecolor{purple}{rgb}{0.35,0,0.35}
\definecolor{orange}{rgb}{1,0.5,0}
\definecolor{darkred}{rgb}{.7,0,0}
\definecolor{darkblue}{rgb}{0,0,.3}
\definecolor{grey}{rgb}{.6,.6,.6}
\definecolor{dimgreen}{rgb}{0.2,0.6,0.1}

\definecolor{darkgreen}{rgb}{0,0.5,0}

\begin{document}

\newcommand{\jav}[1]{{\color{red}#1}}

\title{Work statistics and generalized Loschmidt echo for the Hatano-Nelson model}

\author{Bal\'azs D\'ora}
\email{dora.balazs@ttk.bme.hu}
\affiliation{Department of Theoretical Physics, Institute of Physics, Budapest University of Technology and Economics, M\H uegyetem rkp. 3., H-1111
Budapest, Hungary}
\author{C\u{a}t\u{a}lin Pa\c{s}cu Moca}
\affiliation{MTA-BME Quantum Dynamics and Correlations Research Group, Institute of Physics, Budapest University of Technology and Economics, M\H uegyetem rkp. 3., H-1111, Budapest, Hungary}
\affiliation{Department  of  Physics,  University  of  Oradea,  410087,  Oradea,  Romania}

\date{\today}

\begin{abstract}
We focus on the biorthogonal work statistics of the interacting many-body Hatano-Nelson model after switching on the imaginary vector potential.
We introduce a generalized Loschmidt echo $G(t)$ utilizing the biorthogonal metric operator. It is well suited for numerical analysis and its  
Fourier transform yields the probability distribution 
of work done.
The statistics of work displays several universal features, 
including an exponential decay with the square of both the system size and imaginary vector potential for the probability
to stay in the ground state. 
Additionally, its high energy tail follows  a  universal power law  with exponent $-3$. 
This originates from the peculiar temporal power law decay of $G(t)$ with a time dependent exponent.
The mean and the variance of work scale  linearly and logarithmically with system size  while all higher cumulants are non-extensive.
Our results are relevant for non-unitary field theories as well.
\end{abstract}

\maketitle

\paragraph{Introduction.}
Non-hermitian physics has attracted significant attention recently from almost all  branches of physics, ranging from photonics through
mechanics, active matter and electrical circuits to cold atomic systems and condensed matter\cite{ashidareview,Bergholtz2021}. These
originate from the plethora of unconventional phenomena characteristic to non-hermiticity, including exceptional points\cite{heiss,hodaei,ding2022},
PT-symmetry breaking~\cite{ElGanainy2018},  unidirectional invisibility, non-reciprocal
energy transfer, tachyonic physics\cite{Lee2015}, non-hermitian skin effect, enhanced sensitivity\cite{mcdonald} as well as  non-trivial topological 
phenomena\cite{rotter,gao2015,zhou18,zeuner,gongprx,lee2016,takasu,fruchart,turkeshi,legal,lee2022,kunst2018,gongprx,iscience2019,Lado}.

A significant part of these developments is devoted to biorthogonal and PT-symmetric quantum mechanics\cite{Brody_2014}, which extends hermitian quantum mechanics
by using the distinct left and right eigenstates of non-hermitian Hamiltonians. This initiative is closely connected to  
the investigation of non-unitary  field theories\cite{leeyang,uzelac1979,fisher1978,chang20,sanno,lee2022}.
Given that the
energy represents the most obvious  observable in PT-symmetric biorthogonal systems\cite{meden},
it is natural to investigate quantities depending only on energy. This suggests focusing on thermodynamics and its  
generalizations  non-equilibrium situations\cite{rmptalkner,batalhao} within the non-hermitian realm. 
The celebrated Jarzynski equality\cite{jarzynski} and Crooks relation\cite{crooks} have already been 
extended\cite{gardas,bobowei,deffner2015,Zeng_2017} to biorthogonal quantum systems.
Nevertheless, most of these biorthogonal studies on quantum work  focused on the development of the general theory and its application to simple quantum systems.

Here we take a different approach and study biorthogonal work statistics of one of the most paradigmatic non-hermitian system, 
 the many-body Hatano-Nelson model~\cite{hatanonelson2, hatanonelson1}.
Biorthogonal realizations of this and closely related models have been outlined in Refs. \cite{mcdonald,brandenbourger,mcdonaldprx,kaiwang,hofmann,liu2022,wu2023,wangnc}.
We show that the work statistics in biorthogonal quantum systems can be related to a generalized Loschmidt echo\cite{silva,peres,physrep,goussev}, analogously 
to hermitian systems. By combining bosonization and numerics, we uncover several universal features of the Loschmidt echo and the concomitant work
statistics. We find that the generalized Loschmidt echo, $G(t)$ satisfies a $\ln(G(t))\sim (hvt)^2\ln(vt/L)$
temporal decay with $v$, $h$ and $L$ being the sound velocity, the imaginary vector potential  and the system size. This results in a universal power law scaling
of the probability distribution of quantum work with exponent $-3$. 
Repulsive or attractive interactions renormalize $h$ towards smaller or larger values, respectively. 
With increasing $Lh$, the work distribution crosses over a Gaussian with mean and variance
scaling as $vLh^2$ and $v^2h^2\ln(L)$, respectively.
Our work represents an important step towards analyzing and understanding quantum work in biorthogonal PT-symmetric quantum systems and non-unitary
field theories.

\paragraph{Work statistics and generalized Loschmidt echo.}
For a non-hermitian PT-symmetric Hamiltonian, the right and left eigenstates are defined from\cite{Brody_2014,herviou}
\begin{gather}
H|R_n\rangle =E_n|R_n\rangle,\hspace*{4mm}
H^\dagger|L_n\rangle =E_n|L_n\rangle,
\end{gather}
where $E_n$ is real due to PT-symmetry, and the biorthogonal basis satisfies $\langle L_m|R_n\rangle=\delta_{n,m}$, and the quantum metric operator
is\cite{Brody_2014,gardas}
\begin{gather}
G=\sum_n|L_n\rangle\langle L_n|.
\label{metric}
\end{gather}
Using the metric, the left and right eigenstates are related by $|L_n\rangle=G|R_n\rangle$.
Such PT-symmetric Hamiltonian can be mapped onto a hermitian Hamiltonian $H_h$ with identical spectrum $E_n$  
by a similarity transformation $S$ as
\begin{gather}
H_h=SHS^{-1}
\end{gather}
The metric operator follows from $G=S^\dagger S$. The proper metric\cite{gardas} operator brings it to hermitian diagonal with an additional unitary transformation, 
$UH_hU^\dagger$,
changing $S\rightarrow US$. However, $G$ remains unchanged by this diagonalization.

We consider a system prepared in the ground state $|0\rangle$ of a hermitian Hamiltonian $H_0$ with ground state energy $E_{gs}$. 
Then, at $t=0$ through a sudden quantum quench\cite{polkovnikovrmp,dziarmagareview}, the time evolution
of the system is governed by a non-hermitian PT-symmetric Hamiltonian $H$, which is time independent.
The probability distribution of work\cite{rmptalkner} performed by this non-hermitian process is\cite{gardas,bobowei,deffner2015}
\begin{gather}
P(W)=\sum_n p_n\delta(W-W_n),\hspace*{2mm} p_n=\frac{|\langle L_n|0\rangle|^2}{ \sum_m |\langle L_m|0\rangle|^2},
\end{gather}
where $W_n=E_n-E_{gs}$. While the explicit form of $P(W)$ is very suggestive, its evaluation for a many-body system is very demanding
as it requires the explicit knowledge of several many-body eigenstates.

To circumvent this problem, we define a generalized Loschmidt echo as
\begin{gather}
G(t)=\frac{\langle 0|G\exp(iHt)\exp(-iH_0t)|0\rangle}{\langle 0|G|0\rangle},
\label{genle}
\end{gather}
which is the generating function of work, its Fourier transform yields directly the probability distribution function of work done.
This is seen be inserting Eq. \eqref{metric} into Eq. \eqref{genle} and using the properties of the left eigenstates.
This generalizes previous results from the hermitian case\cite{silva} but is also favourable for numerical or analytical approaches as well.
We note that a related  biorthogonal Loschmidt echo has been analyzed for dynamical quantum phase transition\cite{jing,Tang_2022}.
Using the similarity transformation, this can be recast as
\begin{gather}
G(t)=\frac{\langle 0|S^\dagger\exp(iH_ht)S|0\rangle}{\langle 0|S^\dagger S|0\rangle}\exp(-iE_{gs}t),
\label{genle2}
\end{gather}
which is of the form of the conventional Loschmidt echo for hermitian systems\cite{peres,physrep,goussev}, 
starting from the initial state $S|0\rangle/\sqrt{\langle 0|S^\dagger S|0\rangle}$
and time evolving with the hermitian $H_h$.
This can be useful for simulating the work statistics of non-hermitian systems based on hermitian dynamics.
By implementing a hermitian parent state $H_h$ and a similarity transformation $S$, the generalized Loschmidt echo yields directly the work statistics of non-hermitian
model $S^{-1}H_hS$.

\paragraph{Hatano-Nelson model.}
The Hatano-Nelson model~\cite{hatanonelson2,hatanonelson1} consists of fermions hopping in one dimension in the presence of an imaginary vector potential. Its
interacting many-body Hamiltonian is
\begin{gather}
H_{HN}=\sum_{n=1}^{N-1} \frac J2 \exp(ah)c^\dagger_nc_{n+1}+\frac J2 \exp(-ah) c^\dagger_{n+1}c_n+\nonumber \\
+U c^\dagger_nc_{n}c^\dagger_{n+1}c_{n+1},
\label{hamiltontb}
\end{gather}
where $J>0$ is the uniform hopping, $h$, $a$ and $U$ are the constant 
imaginary vector potential, the lattice constant and  the nearest-neighbour interaction between particles, respectively,
$N$ is the total number of lattice sites. We consider
open boundary condition (OBC) and half filling ($N/2$ particles). 
The model is PT-symmetric\cite{bender2007} and possesses a real spectrum for OBC\cite{PhysRevB.107.045131,PhysRevB.104.195102,zhang2022, alsallom, lee2020,hamazaki, mu2020, zhangprb2020, wang2022,commphyslee,PhysRevB.106.205147}.
It is important to point our that the left and right eigenstates of the Hatano-Nelson model contain the non-hermitian skin effect and are localized to the opposite
ends of the system. However, when evaluating the many-body particle density in the biorthogonal basis, 
it is homogeneous throughout the system and does not reveal any sign
of the skin effect.

Due to the presence of finite $h$, this Hamiltonian is non-hermitian, but can be made hermitian by the similarity transformation
\begin{gather}
S=\exp\left(ha\sum_{n=1}^N n c^\dagger_nc_{n}\right),
\end{gather}
mapping Eq. \eqref{hamiltontb} onto its $h=0$ version. 
This means that the spectrum of the model remains unchanged throughout the quench. 
The exponent of $S$ contains the center of mass operator.

One can construct the effective low-energy Hamiltonian of Eq. \eqref{hamiltontb}~\cite{giamarchi,cazalillaboson,nersesyan,dorahn}, valid in the continuum limit
 using bosonization as
\begin{gather}
H=v \int_0^L \frac{dx}{2\pi} \left[K(\pi\Pi(x)-ih)^2+\frac 1K (\partial_x\phi(x))^2\right],
\label{hamboson}
\end{gather}
where $\Pi(x)$ and $\phi(x)$ are the dual fields satisfying 
the regular commutation relation \cite{cazalillaboson}, $[\Pi(x),\phi(x')]=i\delta(x-x')$.
Eq. \eqref{hamboson} represent a non-unitary field theory as well\cite{fisher1978,uzelac1979,chang20,lee2022} and Eq. \eqref{hamiltontb} its lattice realization.
The similarity transformation
\begin{gather}
S=\exp\left(-\frac{h}{\pi}\int_0^L\phi(x')dx'\right),
\label{smatrix}
\end{gather}
brings it to hermitian  Luttinger liquid (LL) form as
$H_h=\sum_{q>0}\omega_q b^\dagger_qb_q$,
which is also $H_0$.
Its ground state wavefunction,  $|0\rangle$ is the bosonic vacuum and is also the initial state.
The $\phi(x)$ field is expanded in terms of canonical Bose fields respecting OBC\cite{cazalillaboson} as
$\phi(x)=i\sum_{q>0}\sqrt{\frac{\pi K}{qL}}\sin(qx)\left[b_q-b^\dagger_q\right]$,
where $K>0$ is the LL parameter\cite{giamarchi}, which carries all the non-perturbative effects of interaction 
and $\omega_q=vq$ with $v$ the Fermi velocity in the interacting systems and $q=l\pi/L$ with $l=1,2,3\dots$.
Repulsive or attractive interactions give $K<1$ or $K>1$, respectively.
For Eq. \eqref{hamiltontb} with $h=0$, $K=\pi/2/(\pi-\arccos(U/J))$ and $v=aJ{\pi}{\sqrt{1-(U/J)^2}}/{2\arccos(U/J)}$.
The spectrum, $\omega_q$ remains intact during the quench, only the wavefunctions are altered.

\paragraph{Loschmidt echo for the Hatano-Nelson model.}
We apply Eq. \eqref{genle} in combination with Eq. \eqref{smatrix} to determine the work statistics of the Hatano-Nelson model. 
By separating the creation and annihilation operators in Eq. \eqref{smatrix} into two exponentials using the Baker-Campbell-Hausdorff relation\cite{delft}, the resulting expression is a bosonic coherent state\cite{glauber}, when acting on the initial wavefunction $|0\rangle$.
Its time evolution is determined as 
\begin{gather}
\ln[\left(G(t)\right]
=
\frac{4Kh^2}{\pi L}\sum_{q={l\pi}/{L}, l\textmd{ odd}}\frac{\exp(i\omega_qt)-1}{q^3}.
\label{le1}
\end{gather}
This function is periodic in time with periodicity $2L/v$ due to the finite size of the system, and the real and imaginary part of the right hand side
of Eq. \eqref{le1} are symmetric and antisymmetric with respect to $t=L/v$, respectively.
The interaction renormalizes $h$ by the LL parameter to $h\sqrt K$, thus enhancing or suppressing it for attractive or repulsive interactions, respectively.

\begin{figure}[t!]
\centering
\includegraphics[width=7cm]{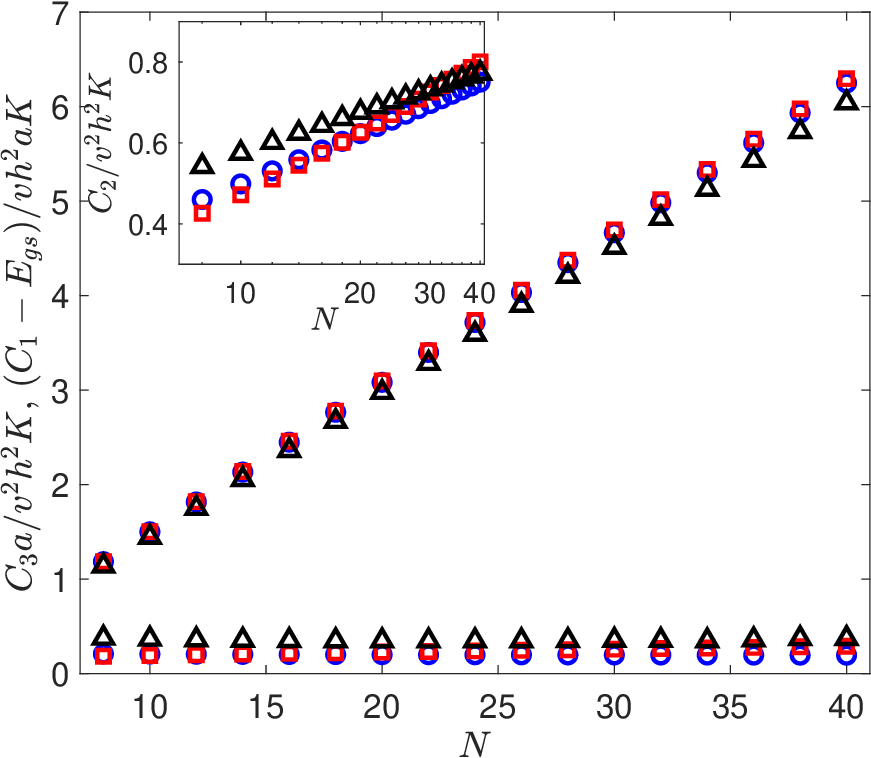}
\caption{The first 3 cumulants of work are shown for several system sizes and for $ha=0.1$, $U/J=0$ (blue circles) and $U/J=0.5$ 
(red squares) as well as for $ha=0.2$, $U/J=-0.5$ (black triangles). The main panel 
depicts the mean (upper symbols) and the 3rd cumulant (lower symbols) while the inset
focuses on the variance of work on a semilogarithmic plot.}
\label{fig:cum}
\end{figure}

The transient, short time dependence of the Loschmidt echo is determined after introducing a high energy cutoff\cite{giamarchi} $\exp(-\alpha |q|)$,
where $\alpha$ is the remnant of the lattice constant $a$ in the continuum limit. 
The short time limit is valid for $t\ll \alpha/v$. Then, the Loschmidt echo is connected to the cumulant generating function of work\cite{silva} by
using the expansion $\ln[\left(G(t)\right]=\sum_{n=1}^\infty C_n (it)^n/n!$.
This yields
\begin{subequations}
\begin{gather}
C_1=\frac{vLh^2K}{2\pi}+E_{gs},\hspace*{3mm}
C_2=\frac{2 (hv)^2K}{\pi^2}\ln\left(\frac{2L}{\pi\alpha}\right),\\
C_{m>2}=\frac{2h^2v^m K (m-3)!}{\pi^2\alpha^{m-2}}.
\end{gather}
\label{cumulants}
\end{subequations}
It is interesting to point out that while the first cumulant is extensive, all higher ones are non-extensive except for
the weak $\ln(L)$ prefactor of the variance of energy.
The cumulants can also be calculated from the moments of energy, which are obtained as $\langle 0|G H^n|0\rangle/\langle 0|G|0\rangle$. 
The Hatano-Nelson model in Eq. \eqref{hamiltontb} is studied by various methods, including many-body exact diagonalization (ED) and matrix product states  techniques~\cite{dmrgmps}
as implemented in Ref.~\cite{itensor}.
The ground state is determined using the density matrix renormalization technique~\cite{White-1992}, and the cumulants are subsequently evaluated numerically, validating the  predictions of Eq. \eqref{cumulants} in Fig. \ref{fig:cum} already for relatively small systems.
In principle, $\alpha$ can also get slightly renormalized by both $U$ and $h$.

\begin{figure}[t!]
\centering
\includegraphics[width=7cm]{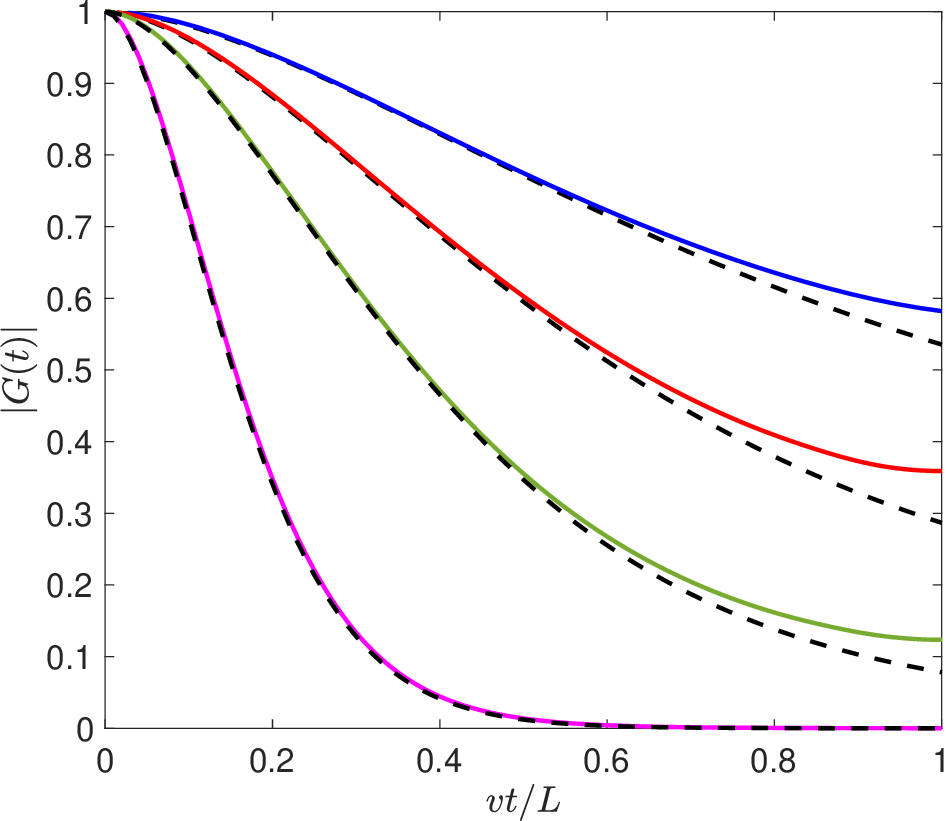}
\caption{The decay of the Loschmidt echo  is shown for $ha=0.1$, $N=28$, $U/J=0.5$ (blue line) and $U/J=-0.5$ (red line), $N=40$, $U/J=-0.5$ (green line) 
as well as for $N=100$, $U=0$ (magenta line)
for Eq. \eqref{hamiltontb}.  
The black dashed lines denote Eq. \eqref{ttot2} without any free parameter, while Eq. \eqref{leexact} is indistinguishable from ED data. }
\label{fig:ledecay}
\end{figure}

For longer times $t\gg\alpha/v$, Eq. \eqref{le1} is evaluated in closed form without the need of using the cutoff as
\begin{gather}
\ln[\left(G(t)\right]=2K\left(\frac{Lh}{\pi^2}\right)^2\times\nonumber\\
\times\left[-\frac{7\zeta(3)}{4}+\sum_{\sigma=\pm}\sigma \textmd{Li}_3[\sigma\exp(ivt\pi/L)]\right],
\label{leexact}
\end{gather}
where Li$_3(z)$ the polylogarithm function\cite{gradstein} of order $3$ and $\zeta(3)\approx 1.202$.
Its most peculiar feature is its temporal decay in the $\alpha\ll vt\ll L$ limit, given by
\begin{gather}
|G(t)|\simeq \left(\frac{\pi vt}{2L\exp(3/2)}\right)^{K(hvt/\pi)^2},
\label{ttot2}
\end{gather}
featuring an unusual $t^{c t^2}$ behaviour in the scaling limit. 
We compare this to numerics in Fig. \ref{fig:ledecay} with perfect agreement already for small systems. In addition to many-body ED for $L\leq 28$, we also performed
single particle ED based on Eq. \eqref{genle2} in the non-interacting limit, reaching system sizes of the order of a few hundred sites.
This was complemented by time evolving block decimation calculations\cite{Vidal-2003} for interacting systems.
As we demonstrate below, Eq. \eqref{ttot2} is responsible for the universal power law decay of the work distribution.

The overall qualitative shape of Eq. \eqref{leexact} is well approximated by the cumulant generating function of the Poisson distribution as
$\ln[\left(G(t)\right]\approx \frac{7\zeta(3)K}{2}\left(\frac{Lh}{\pi^2}\right)^2\left(\exp(ivt\pi/L)-1\right)$,
which obviously does not capture Eq. \eqref{ttot2}.

\paragraph{Biorthogonal work statistics.}
Due to the finite size of the system, the level spacing is finite, thus the work can only take integer multiples of $\frac{v\pi}{L}$.
The associated work distribution is discrete as $P(W=\frac{vn\pi}{L})=\frac{v}{2L}\int_0^{2L/v}dt~ G(t)\exp(-iWt)dt$. While this holds true for the bosonization results,
it is only approximately true for the lattice version of the Hatano-Nelson model. The energy level spacing is not uniform and the revival after $2L/v$ time
is not perfect, therefore integrating to infinity in the Fourier transform would in principle be needed. However, one can argue that by taking the thermodynamic limit
in the system size, the time evolution would hardly reach even the first revival at $2L/v$. In addition, the generalized Loschmidt echo displays almost identical shapes
during the first few revivals thus focusing on the first is a good approximation.

Our results allow us to evaluate the probability of no work done ($P(W=0)$), 
namely the probability to stay in the ground state of the Hatano-Nelson model after the imaginary vector potential quench.
This reads as
\begin{gather}
P_{ad}=\exp\left(-\frac{7\zeta(3)K}{2}\left(\frac{Lh}{\pi^2}\right)^2\right),
\label{pad}
\end{gather}
and gets suppressed with $K(Lh)^2$.

\begin{figure}[t!]
\centering
\includegraphics[width=7cm]{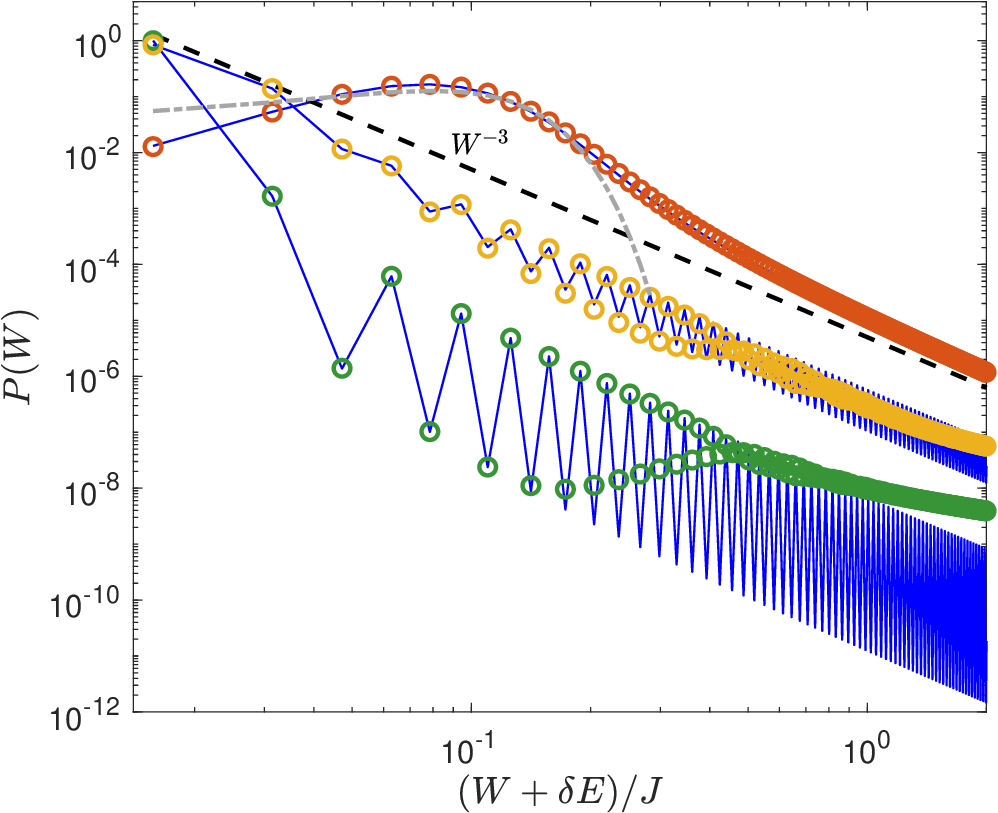}
\caption{The work distribution is plotted in the non-interacting limit for $N=200$ and $ha=0.05$ (red circles), 0.01 (orange circles) and  0.001 (green circles) 
from to to bottom from
the full numerical solution of Eq. \eqref{hamiltontb}, $W$ is integer multiple of $v\pi/L$.
The horizontal scale is shifted by the level spacing $\delta E=J\pi/N=v\pi/L$ to make $W=0$ visible on log scale.
The gray dash-dotted line is a  gaussian from Eqs. \eqref{cumulants} for the $ha=0.05$ data.
The blue solid lines represent the numerical Fourier transform of Eq. \eqref{leexact} at integer multiple of $v\pi/L$ and is a guide to the eye in between data points.
The black dashed line denotes $1/W^3$ asymptote.}
\label{fig:pw}
\end{figure}

Upon increasing the energy, the distribution function can develop a peak at around the mean energy, and decays afterwards as $W^{-3}$ in a universal fashion.
Then, additional high energy features arising from band curvature and the underlying lattice can kick in, which is beyond the validity of bosonization.
This universal power law decay can be understood already from the small system or small imaginary vector potential limit ($Lh\ll1$).
The generalized Loschmidt echo, $G(t)$ in Eq. \eqref{leexact} is Taylor expanded to $h^2$ order, allowing for an analytic Fourier transform.
The probability of staying in the ground state still follows from the small $h$ limit of Eq. \eqref{pad} as 
$P_{ad}=1-\frac{7\zeta(3)K}{2}\left(\frac{Lh}{\pi^2}\right)^2$.
The probability of finite work done is
\begin{gather}
P\left(W=(2l+1)\frac{v\pi}{L}\right)=\frac{4Kh^2v^3}{ \pi L}\times W^{-3}
\label{pwpert}
\end{gather}
with $l$ non-negative integer. By including higher order terms, $P(W= 2l{v\pi}/{L})\sim v^3 h^4 L W^{-3}$
 and $P(W)$ is zero otherwise. This predicts not only a power law decay but also  oscillating behaviour between even or odd multiples of the level spacing.
This universal power law decay remains present for large $Lh$ and is a direct consequence of Eq. \eqref{ttot2}, 
as follows from a numerical Fourier transform of Eq. \eqref{leexact}.
We use Eq. \eqref{genle2} in combination with single particle ED to evaluate the generalized Loschmidt echo, what is subsequently Fourier 
transformed numerically,  shown in Fig. \ref{fig:pw}.
While bosonization is an effective low energy method, it works surprisingly well up to energies comparable to the single particle bandwidth, $J$, deviations between
the continuum and lattice calculations only start to appear for larger energies~\footnote{This can be an artifact of the finite temporal window of Fourier transform 
and the non-perfect
periodic behaviour of the generalized Loschmidt echo of the lattice model due to high energy modes. This is corroborated by the fact that while the $P(W<J)$ is robust
with respect to variations of the temporal window of Fourier transform, the high energy part in Fig. \ref{fig:pw} is very sensitive 
to even small variations of the time period.}.

With increasing $Lh$, the work distribution smoothly crosses over to a gaussian sharp peak as expected from the scaling of the cumulants in Eqs. \eqref{cumulants}, whose formation is visualized in Fig. \ref{fig:pw} by the gray curve. Its high energy tail still retains
the universal power law form. Thus, the universal low energy features are perfectly captured by bosonization, even for a sudden, non-hermitian quench.

\paragraph{Conclusions.}
We introduced a generalized Loschmidt echo tailored for biorthogonal, PT-symmetric non-hermitian quantum system.
This establishes a direct connection with work statistics
from a non-equilibrium process. This framework was effectively employed for  the interacting Hatano-Nelson model after a sudden quench of the imaginary vector potential.
Repulsive or attractive interactions renormalize the imaginary vector potential to smaller or bigger values, respectively. 
We analyzed the evolution of work statistics and identified several universal features, such as the power law decay of work with exponent $-3$ and
the scaling of the cumulants of work with system size and non-hermitian parameter. 
Our work also contributes to extending work statistics to non-unitary field theories\cite{fisher1978}.

\begin{acknowledgments}
This research is supported by the National Research, Development and 
Innovation Office - NKFIH  within the Quantum Technology National Excellence 
Program (Project No.~2017-1.2.1-NKP-2017-00001), K134437, K142179 by the BME-Nanotechnology 
FIKP grant (BME FIKP-NAT), and by a grant of the Ministry of Research, Innovation and
 Digitization, CNCS/CCCDI-UEFISCDI, under projects number PN-III-P4-ID-PCE-2020-0277 and 
 under the project for funding the excellence, Contract No. 29 PFE/30.12.2021.
\end{acknowledgments}

\bibliographystyle{apsrev}
\bibliography{wboson1}

\begin{thebibliography}{10}
\expandafter\ifx\csname bibnamefont\endcsname\relax
  \def\bibnamefont#1{#1}\fi
\expandafter\ifx\csname bibfnamefont\endcsname\relax
  \def\bibfnamefont#1{#1}\fi
\expandafter\ifx\csname url\endcsname\relax
  \def\url#1{\texttt{#1}}\fi
\expandafter\ifx\csname urlprefix\endcsname\relax\def\urlprefix{URL }\fi
\providecommand{\bibinfo}[2]{#2}
\providecommand{\eprint}[2][]{\url{#2}}

\bibitem{ashidareview}
\bibinfo{author}{\bibfnamefont{Y.}~\bibnamefont{Ashida}},
  \bibinfo{author}{\bibfnamefont{Z.}~\bibnamefont{Gong}}, \bibnamefont{and}
  \bibinfo{author}{\bibfnamefont{M.}~\bibnamefont{Ueda}},
  \emph{\bibinfo{title}{Non-hermitian physics}}, \bibinfo{journal}{Advances in
  Physics} \textbf{\bibinfo{volume}{69}}, \bibinfo{pages}{3}
  (\bibinfo{year}{2020}).

\bibitem{Bergholtz2021}
\bibinfo{author}{\bibfnamefont{E.~J.} \bibnamefont{Bergholtz}},
  \bibinfo{author}{\bibfnamefont{J.~C.} \bibnamefont{Budich}},
  \bibnamefont{and} \bibinfo{author}{\bibfnamefont{F.~K.} \bibnamefont{Kunst}},
  \emph{\bibinfo{title}{Exceptional topology of non-hermitian systems}},
  \bibinfo{journal}{Rev. Mod. Phys.} \textbf{\bibinfo{volume}{93}},
  \bibinfo{pages}{015005} (\bibinfo{year}{2021}).

\bibitem{heiss}
\bibinfo{author}{\bibfnamefont{W.~D.} \bibnamefont{Heiss}},
  \emph{\bibinfo{title}{The physics of exceptional points}},
  \bibinfo{journal}{J. Phys. A: Math. Theor.}
  \textbf{\bibinfo{volume}{45}}(\bibinfo{number}{44}), \bibinfo{pages}{444016}
  (\bibinfo{year}{2012}).

\bibitem{hodaei}
\bibinfo{author}{\bibfnamefont{H.}~\bibnamefont{Hodaei}},
  \bibinfo{author}{\bibfnamefont{A.~U.} \bibnamefont{Hassan}},
  \bibinfo{author}{\bibfnamefont{S.}~\bibnamefont{Wittek}},
  \bibinfo{author}{\bibfnamefont{H.}~\bibnamefont{Garcia-Gracia}},
  \bibinfo{author}{\bibfnamefont{R.}~\bibnamefont{El-Ganainy}},
  \bibinfo{author}{\bibfnamefont{D.~N.} \bibnamefont{Christodoulides}},
  \bibnamefont{and}
  \bibinfo{author}{\bibfnamefont{M.}~\bibnamefont{Khajavikhan}},
  \emph{\bibinfo{title}{Enhanced sensitivity at higher-order exceptional
  points}}, \bibinfo{journal}{Nature} \textbf{\bibinfo{volume}{548}},
  \bibinfo{pages}{187} (\bibinfo{year}{2017}).

\bibitem{ding2022}
\bibinfo{author}{\bibfnamefont{K.}~\bibnamefont{Ding}},
  \bibinfo{author}{\bibfnamefont{C.}~\bibnamefont{Fang}}, \bibnamefont{and}
  \bibinfo{author}{\bibfnamefont{G.}~\bibnamefont{Ma}},
  \emph{\bibinfo{title}{Non-hermitian topology and exceptional-point
  geometries}}, \bibinfo{journal}{Nature Reviews Physics}
  \textbf{\bibinfo{volume}{4}}, \bibinfo{pages}{745} (\bibinfo{year}{2022}).

\bibitem{ElGanainy2018}
\bibinfo{author}{\bibfnamefont{R.}~\bibnamefont{El-Ganainy}},
  \bibinfo{author}{\bibfnamefont{K.~G.} \bibnamefont{Makris}},
  \bibinfo{author}{\bibfnamefont{M.}~\bibnamefont{Khajavikhan}},
  \bibinfo{author}{\bibfnamefont{Z.~H.} \bibnamefont{Musslimani}},
  \bibinfo{author}{\bibfnamefont{S.}~\bibnamefont{Rotter}}, \bibnamefont{and}
  \bibinfo{author}{\bibfnamefont{D.~N.} \bibnamefont{Christodoulides}},
  \emph{\bibinfo{title}{Non-hermitian physics and pt symmetry}},
  \bibinfo{journal}{Nat. Phys.}
  \textbf{\bibinfo{volume}{14}}(\bibinfo{number}{1}), \bibinfo{pages}{11}
  (\bibinfo{year}{2018}).

\bibitem{Lee2015}
\bibinfo{author}{\bibfnamefont{T.~E.} \bibnamefont{Lee}},
  \bibinfo{author}{\bibfnamefont{U.}~\bibnamefont{Alvarez-Rodriguez}},
  \bibinfo{author}{\bibfnamefont{X.-H.} \bibnamefont{Cheng}},
  \bibinfo{author}{\bibfnamefont{L.}~\bibnamefont{Lamata}}, \bibnamefont{and}
  \bibinfo{author}{\bibfnamefont{E.}~\bibnamefont{Solano}},
  \emph{\bibinfo{title}{Tachyon physics with trapped ions}},
  \bibinfo{journal}{Phys. Rev. A} \textbf{\bibinfo{volume}{92}},
  \bibinfo{pages}{032129} (\bibinfo{year}{2015}).

\bibitem{mcdonald}
\bibinfo{author}{\bibfnamefont{A.}~\bibnamefont{McDonald}} \bibnamefont{and}
  \bibinfo{author}{\bibfnamefont{A.~A.} \bibnamefont{Clerk}},
  \emph{\bibinfo{title}{Exponentially-enhanced quantum sensing with
  non-hermitian lattice dynamics}}, \bibinfo{journal}{Nat. Commun.}
  \textbf{\bibinfo{volume}{11}}, \bibinfo{pages}{5382} (\bibinfo{year}{2020}).

\bibitem{rotter}
\bibinfo{author}{\bibfnamefont{I.}~\bibnamefont{Rotter}} \bibnamefont{and}
  \bibinfo{author}{\bibfnamefont{J.~P.} \bibnamefont{Bird}},
  \emph{\bibinfo{title}{A review of progress in the physics of open quantum
  systems: theory and experiment}}, \bibinfo{journal}{Rep. Prog. Phys.}
  \textbf{\bibinfo{volume}{78}}, \bibinfo{pages}{114001}
  (\bibinfo{year}{2015}).

\bibitem{gao2015}
\bibinfo{author}{\bibfnamefont{T.}~\bibnamefont{Gao}},
  \bibinfo{author}{\bibfnamefont{E.}~\bibnamefont{Estrecho}},
  \bibinfo{author}{\bibfnamefont{K.~Y.} \bibnamefont{Bliokh}},
  \bibinfo{author}{\bibfnamefont{T.~C.~H.} \bibnamefont{Liew}},
  \bibinfo{author}{\bibfnamefont{M.~D.} \bibnamefont{Fraser}},
  \bibinfo{author}{\bibfnamefont{S.}~\bibnamefont{Brodbeck}},
  \bibinfo{author}{\bibfnamefont{M.}~\bibnamefont{Kamp}},
  \bibinfo{author}{\bibfnamefont{C.}~\bibnamefont{Schneider}},
  \bibinfo{author}{\bibfnamefont{S.}~\bibnamefont{H{\"o}fling}},
  \bibinfo{author}{\bibfnamefont{Y.}~\bibnamefont{Yamamoto}},
  \bibinfo{author}{\bibfnamefont{F.}~\bibnamefont{Nori}},
  \bibinfo{author}{\bibfnamefont{Y.~S.} \bibnamefont{Kivshar}}, \emph{et~al.},
  \emph{\bibinfo{title}{Observation of non-hermitian degeneracies in a chaotic
  exciton-polariton billiard}}, \bibinfo{journal}{Nature}
  \textbf{\bibinfo{volume}{526}}, \bibinfo{pages}{554} (\bibinfo{year}{2015}).

\bibitem{zhou18}
\bibinfo{author}{\bibfnamefont{L.}~\bibnamefont{Zhou}},
  \bibinfo{author}{\bibfnamefont{Q.-h.} \bibnamefont{Wang}},
  \bibinfo{author}{\bibfnamefont{H.}~\bibnamefont{Wang}}, \bibnamefont{and}
  \bibinfo{author}{\bibfnamefont{J.}~\bibnamefont{Gong}},
  \emph{\bibinfo{title}{Dynamical quantum phase transitions in non-hermitian
  lattices}}, \bibinfo{journal}{Phys. Rev. A} \textbf{\bibinfo{volume}{98}},
  \bibinfo{pages}{022129} (\bibinfo{year}{2018}).

\bibitem{zeuner}
\bibinfo{author}{\bibfnamefont{J.~M.} \bibnamefont{Zeuner}},
  \bibinfo{author}{\bibfnamefont{M.~C.} \bibnamefont{Rechtsman}},
  \bibinfo{author}{\bibfnamefont{Y.}~\bibnamefont{Plotnik}},
  \bibinfo{author}{\bibfnamefont{Y.}~\bibnamefont{Lumer}},
  \bibinfo{author}{\bibfnamefont{S.}~\bibnamefont{Nolte}},
  \bibinfo{author}{\bibfnamefont{M.~S.} \bibnamefont{Rudner}},
  \bibinfo{author}{\bibfnamefont{M.}~\bibnamefont{Segev}}, \bibnamefont{and}
  \bibinfo{author}{\bibfnamefont{A.}~\bibnamefont{Szameit}},
  \emph{\bibinfo{title}{Observation of a topological transition in the bulk of
  a non-hermitian system}}, \bibinfo{journal}{Phys. Rev. Lett.}
  \textbf{\bibinfo{volume}{115}}, \bibinfo{pages}{040402}
  (\bibinfo{year}{2015}).

\bibitem{gongprx}
\bibinfo{author}{\bibfnamefont{Z.}~\bibnamefont{Gong}},
  \bibinfo{author}{\bibfnamefont{Y.}~\bibnamefont{Ashida}},
  \bibinfo{author}{\bibfnamefont{K.}~\bibnamefont{Kawabata}},
  \bibinfo{author}{\bibfnamefont{K.}~\bibnamefont{Takasan}},
  \bibinfo{author}{\bibfnamefont{S.}~\bibnamefont{Higashikawa}},
  \bibnamefont{and} \bibinfo{author}{\bibfnamefont{M.}~\bibnamefont{Ueda}},
  \emph{\bibinfo{title}{Topological phases of non-hermitian systems}},
  \bibinfo{journal}{Phys. Rev. X} \textbf{\bibinfo{volume}{8}},
  \bibinfo{pages}{031079} (\bibinfo{year}{2018}).

\bibitem{lee2016}
\bibinfo{author}{\bibfnamefont{T.~E.} \bibnamefont{Lee}},
  \emph{\bibinfo{title}{Anomalous edge state in a non-hermitian lattice}},
  \bibinfo{journal}{Phys. Rev. Lett.} \textbf{\bibinfo{volume}{116}},
  \bibinfo{pages}{133903} (\bibinfo{year}{2016}).

\bibitem{takasu}
\bibinfo{author}{\bibfnamefont{Y.}~\bibnamefont{Takasu}},
  \bibinfo{author}{\bibfnamefont{T.}~\bibnamefont{Yagami}},
  \bibinfo{author}{\bibfnamefont{Y.}~\bibnamefont{Ashida}},
  \bibinfo{author}{\bibfnamefont{R.}~\bibnamefont{Hamazaki}},
  \bibinfo{author}{\bibfnamefont{Y.}~\bibnamefont{Kuno}}, \bibnamefont{and}
  \bibinfo{author}{\bibfnamefont{Y.}~\bibnamefont{Takahashi}},
  \emph{\bibinfo{title}{{PT-symmetric non-Hermitian quantum many-body system
  using ultracold atoms in an optical lattice with controlled dissipation}}},
  \bibinfo{journal}{Progress of Theoretical and Experimental Physics}
  \textbf{\bibinfo{volume}{2020}}(\bibinfo{number}{12}) (\bibinfo{year}{2020}),
  \bibinfo{note}{12A110}.

\bibitem{fruchart}
\bibinfo{author}{\bibfnamefont{M.}~\bibnamefont{Fruchart}},
  \bibinfo{author}{\bibfnamefont{R.}~\bibnamefont{Hanai}},
  \bibinfo{author}{\bibfnamefont{P.~B.} \bibnamefont{Littlewood}},
  \bibnamefont{and} \bibinfo{author}{\bibfnamefont{V.}~\bibnamefont{Vitelli}},
  \emph{\bibinfo{title}{Non-reciprocal phase transitions}},
  \bibinfo{journal}{Nature} \textbf{\bibinfo{volume}{592}},
  \bibinfo{pages}{363} (\bibinfo{year}{2021}).

\bibitem{turkeshi}
\bibinfo{author}{\bibfnamefont{X.}~\bibnamefont{Turkeshi}} \bibnamefont{and}
  \bibinfo{author}{\bibfnamefont{M.}~\bibnamefont{Schir\'o}},
  \emph{\bibinfo{title}{Entanglement and correlation spreading in non-hermitian
  spin chains}}, \bibinfo{journal}{Phys. Rev. B}
  \textbf{\bibinfo{volume}{107}}, \bibinfo{pages}{L020403}
  (\bibinfo{year}{2023}).

\bibitem{legal}
\bibinfo{author}{\bibfnamefont{Y.~L.} \bibnamefont{Gal}},
  \bibinfo{author}{\bibfnamefont{X.}~\bibnamefont{Turkeshi}}, \bibnamefont{and}
  \bibinfo{author}{\bibfnamefont{M.}~\bibnamefont{Schirò}},
  \emph{\bibinfo{title}{{Volume-to-area law entanglement transition in a
  non-Hermitian free fermionic chain}}}, \bibinfo{journal}{SciPost Phys.}
  \textbf{\bibinfo{volume}{14}}, \bibinfo{pages}{138} (\bibinfo{year}{2023}).

\bibitem{lee2022}
\bibinfo{author}{\bibfnamefont{C.~H.} \bibnamefont{Lee}},
  \emph{\bibinfo{title}{Exceptional bound states and negative entanglement
  entropy}}, \bibinfo{journal}{Phys. Rev. Lett.}
  \textbf{\bibinfo{volume}{128}}, \bibinfo{pages}{010402}
  (\bibinfo{year}{2022}).

\bibitem{kunst2018}
\bibinfo{author}{\bibfnamefont{F.~K.} \bibnamefont{Kunst}},
  \bibinfo{author}{\bibfnamefont{E.}~\bibnamefont{Edvardsson}},
  \bibinfo{author}{\bibfnamefont{J.~C.} \bibnamefont{Budich}},
  \bibnamefont{and} \bibinfo{author}{\bibfnamefont{E.~J.}
  \bibnamefont{Bergholtz}}, \emph{\bibinfo{title}{Biorthogonal bulk-boundary
  correspondence in non-hermitian systems}}, \bibinfo{journal}{Phys. Rev.
  Lett.} \textbf{\bibinfo{volume}{121}}, \bibinfo{pages}{026808}
  (\bibinfo{year}{2018}).

\bibitem{iscience2019}
\bibinfo{author}{\bibfnamefont{X.}~\bibnamefont{Qiu}},
  \bibinfo{author}{\bibfnamefont{T.-S.} \bibnamefont{Deng}},
  \bibinfo{author}{\bibfnamefont{Y.}~\bibnamefont{Hu}},
  \bibinfo{author}{\bibfnamefont{P.}~\bibnamefont{Xue}}, \bibnamefont{and}
  \bibinfo{author}{\bibfnamefont{W.}~\bibnamefont{Yi}},
  \emph{\bibinfo{title}{Fixed points and dynamic topological phenomena in a
  parity-time-symmetric quantum quench}}, \bibinfo{journal}{iScience}
  \textbf{\bibinfo{volume}{20}}, \bibinfo{pages}{392} (\bibinfo{year}{2019}).

\bibitem{Lado}
\bibinfo{author}{\bibfnamefont{G.}~\bibnamefont{Chen}},
  \bibinfo{author}{\bibfnamefont{F.}~\bibnamefont{Song}}, \bibnamefont{and}
  \bibinfo{author}{\bibfnamefont{J.~L.} \bibnamefont{Lado}},
  \emph{\bibinfo{title}{Topological spin excitations in non-hermitian spin
  chains with a generalized kernel polynomial algorithm}},
  \bibinfo{journal}{Phys. Rev. Lett.} \textbf{\bibinfo{volume}{130}},
  \bibinfo{pages}{100401} (\bibinfo{year}{2023}).

\bibitem{Brody_2014}
\bibinfo{author}{\bibfnamefont{D.~C.} \bibnamefont{Brody}},
  \emph{\bibinfo{title}{Biorthogonal quantum mechanics}},
  \bibinfo{journal}{Journal of Physics A: Mathematical and Theoretical}
  \textbf{\bibinfo{volume}{47}}(\bibinfo{number}{3}), \bibinfo{pages}{035305}
  (\bibinfo{year}{2013}).

\bibitem{leeyang}
\bibinfo{author}{\bibfnamefont{T.~D.} \bibnamefont{Lee}} \bibnamefont{and}
  \bibinfo{author}{\bibfnamefont{C.~N.} \bibnamefont{Yang}},
  \emph{\bibinfo{title}{Statistical theory of equations of state and phase
  transitions. ii. lattice gas and ising model}}, \bibinfo{journal}{Phys. Rev.}
  \textbf{\bibinfo{volume}{87}}, \bibinfo{pages}{410} (\bibinfo{year}{1952}).

\bibitem{uzelac1979}
\bibinfo{author}{\bibfnamefont{K.}~\bibnamefont{Uzelac}},
  \bibinfo{author}{\bibfnamefont{P.}~\bibnamefont{Pfeuty}}, \bibnamefont{and}
  \bibinfo{author}{\bibfnamefont{R.}~\bibnamefont{Jullien}},
  \emph{\bibinfo{title}{Yang-lee edge singularity from a real-space
  renormalization-group method}}, \bibinfo{journal}{Phys. Rev. Lett.}
  \textbf{\bibinfo{volume}{43}}, \bibinfo{pages}{805} (\bibinfo{year}{1979}).

\bibitem{fisher1978}
\bibinfo{author}{\bibfnamefont{M.~E.} \bibnamefont{Fisher}},
  \emph{\bibinfo{title}{Yang-lee edge singularity and ${\ensuremath{\phi}}^{3}$
  field theory}}, \bibinfo{journal}{Phys. Rev. Lett.}
  \textbf{\bibinfo{volume}{40}}, \bibinfo{pages}{1610} (\bibinfo{year}{1978}).

\bibitem{chang20}
\bibinfo{author}{\bibfnamefont{P.-Y.} \bibnamefont{Chang}},
  \bibinfo{author}{\bibfnamefont{J.-S.} \bibnamefont{You}},
  \bibinfo{author}{\bibfnamefont{X.}~\bibnamefont{Wen}}, \bibnamefont{and}
  \bibinfo{author}{\bibfnamefont{S.}~\bibnamefont{Ryu}},
  \emph{\bibinfo{title}{Entanglement spectrum and entropy in topological
  non-hermitian systems and nonunitary conformal field theory}},
  \bibinfo{journal}{Phys. Rev. Research} \textbf{\bibinfo{volume}{2}},
  \bibinfo{pages}{033069} (\bibinfo{year}{2020}).

\bibitem{sanno}
\bibinfo{author}{\bibfnamefont{T.}~\bibnamefont{Sanno}},
  \bibinfo{author}{\bibfnamefont{M.~G.} \bibnamefont{Yamada}},
  \bibinfo{author}{\bibfnamefont{T.}~\bibnamefont{Mizushima}},
  \bibnamefont{and} \bibinfo{author}{\bibfnamefont{S.}~\bibnamefont{Fujimoto}},
  \emph{\bibinfo{title}{Engineering yang-lee anyons via majorana bound
  states}}, \bibinfo{journal}{Phys. Rev. B} \textbf{\bibinfo{volume}{106}},
  \bibinfo{pages}{174517} (\bibinfo{year}{2022}).

\bibitem{meden}
\bibinfo{author}{\bibfnamefont{V.}~\bibnamefont{Meden}},
  \bibinfo{author}{\bibfnamefont{L.}~\bibnamefont{Grunwald}}, \bibnamefont{and}
  \bibinfo{author}{\bibfnamefont{D.~M.} \bibnamefont{Kennes}},
  \emph{\bibinfo{title}{Pt-symmetric, non-hermitian quantum many-body physics
  -- a methodological perspective}}, \bibinfo{note}{{a}rXiv:2303.05956}.

\bibitem{rmptalkner}
\bibinfo{author}{\bibfnamefont{M.}~\bibnamefont{Campisi}},
  \bibinfo{author}{\bibfnamefont{P.}~\bibnamefont{H\"anggi}}, \bibnamefont{and}
  \bibinfo{author}{\bibfnamefont{P.}~\bibnamefont{Talkner}},
  \emph{\bibinfo{title}{\textit{Colloquium} : Quantum fluctuation relations:
  Foundations and applications}}, \bibinfo{journal}{Rev. Mod. Phys.}
  \textbf{\bibinfo{volume}{83}}, \bibinfo{pages}{771} (\bibinfo{year}{2011}).

\bibitem{batalhao}
\bibinfo{author}{\bibfnamefont{T.~B.} \bibnamefont{Batalh\~ao}},
  \bibinfo{author}{\bibfnamefont{A.~M.} \bibnamefont{Souza}},
  \bibinfo{author}{\bibfnamefont{L.}~\bibnamefont{Mazzola}},
  \bibinfo{author}{\bibfnamefont{R.}~\bibnamefont{Auccaise}},
  \bibinfo{author}{\bibfnamefont{R.~S.} \bibnamefont{Sarthour}},
  \bibinfo{author}{\bibfnamefont{I.~S.} \bibnamefont{Oliveira}},
  \bibinfo{author}{\bibfnamefont{J.}~\bibnamefont{Goold}},
  \bibinfo{author}{\bibfnamefont{G.}~\bibnamefont{De~Chiara}},
  \bibinfo{author}{\bibfnamefont{M.}~\bibnamefont{Paternostro}},
  \bibnamefont{and} \bibinfo{author}{\bibfnamefont{R.~M.} \bibnamefont{Serra}},
  \emph{\bibinfo{title}{Experimental reconstruction of work distribution and
  study of fluctuation relations in a closed quantum system}},
  \bibinfo{journal}{Phys. Rev. Lett.} \textbf{\bibinfo{volume}{113}},
  \bibinfo{pages}{140601} (\bibinfo{year}{2014}).

\bibitem{jarzynski}
\bibinfo{author}{\bibfnamefont{C.}~\bibnamefont{Jarzynski}},
  \emph{\bibinfo{title}{Nonequilibrium equality for free energy differences}},
  \bibinfo{journal}{Phys. Rev. Lett.} \textbf{\bibinfo{volume}{78}},
  \bibinfo{pages}{2690} (\bibinfo{year}{1997}).

\bibitem{crooks}
\bibinfo{author}{\bibfnamefont{G.~E.} \bibnamefont{Crooks}},
  \emph{\bibinfo{title}{Entropy production fluctuation theorem and the
  nonequilibrium work relation for free energy differences}},
  \bibinfo{journal}{Phys. Rev. E} \textbf{\bibinfo{volume}{60}},
  \bibinfo{pages}{2721} (\bibinfo{year}{1999}).

\bibitem{gardas}
\bibinfo{author}{\bibfnamefont{B.}~\bibnamefont{Gardas}},
  \bibinfo{author}{\bibfnamefont{S.}~\bibnamefont{Deffner}}, \bibnamefont{and}
  \bibinfo{author}{\bibfnamefont{A.}~\bibnamefont{Saxena}},
  \emph{\bibinfo{title}{Non-hermitian quantum thermodynamics}},
  \bibinfo{journal}{Scientific Reports} \textbf{\bibinfo{volume}{6}},
  \bibinfo{pages}{23408} (\bibinfo{year}{2016}).

\bibitem{bobowei}
\bibinfo{author}{\bibfnamefont{B.-B.} \bibnamefont{Wei}},
  \emph{\bibinfo{title}{Quantum work relations and response theory in
  parity-time-symmetric quantum systems}}, \bibinfo{journal}{Phys. Rev. E}
  \textbf{\bibinfo{volume}{97}}, \bibinfo{pages}{012114}
  (\bibinfo{year}{2018}).

\bibitem{deffner2015}
\bibinfo{author}{\bibfnamefont{S.}~\bibnamefont{Deffner}} \bibnamefont{and}
  \bibinfo{author}{\bibfnamefont{A.}~\bibnamefont{Saxena}},
  \emph{\bibinfo{title}{Jarzynski equality in
  $\mathcal{P}\mathcal{T}$-symmetric quantum mechanics}},
  \bibinfo{journal}{Phys. Rev. Lett.} \textbf{\bibinfo{volume}{114}},
  \bibinfo{pages}{150601} (\bibinfo{year}{2015}).

\bibitem{Zeng_2017}
\bibinfo{author}{\bibfnamefont{M.}~\bibnamefont{Zeng}} \bibnamefont{and}
  \bibinfo{author}{\bibfnamefont{E.~H.} \bibnamefont{Yong}},
  \emph{\bibinfo{title}{Crooks fluctuation theorem in {{PT}}-symmetric quantum
  mechanics}}, \bibinfo{journal}{Journal of Physics Communications}
  \textbf{\bibinfo{volume}{1}}(\bibinfo{number}{3}), \bibinfo{pages}{031001}
  (\bibinfo{year}{2017}).

\bibitem{hatanonelson2}
\bibinfo{author}{\bibfnamefont{N.}~\bibnamefont{Hatano}} \bibnamefont{and}
  \bibinfo{author}{\bibfnamefont{D.~R.} \bibnamefont{Nelson}},
  \emph{\bibinfo{title}{Localization transitions in non-hermitian quantum
  mechanics}}, \bibinfo{journal}{Phys. Rev. Lett.}
  \textbf{\bibinfo{volume}{77}}, \bibinfo{pages}{570} (\bibinfo{year}{1996}).

\bibitem{hatanonelson1}
\bibinfo{author}{\bibfnamefont{N.}~\bibnamefont{Hatano}} \bibnamefont{and}
  \bibinfo{author}{\bibfnamefont{D.~R.} \bibnamefont{Nelson}},
  \emph{\bibinfo{title}{Vortex pinning and non-hermitian quantum mechanics}},
  \bibinfo{journal}{Phys. Rev. B} \textbf{\bibinfo{volume}{56}},
  \bibinfo{pages}{8651} (\bibinfo{year}{1997}).

\bibitem{brandenbourger}
\bibinfo{author}{\bibfnamefont{M.}~\bibnamefont{Brandenbourger}},
  \bibinfo{author}{\bibfnamefont{X.}~\bibnamefont{Locsin}},
  \bibinfo{author}{\bibfnamefont{E.}~\bibnamefont{Lerner}}, \bibnamefont{and}
  \bibinfo{author}{\bibfnamefont{C.}~\bibnamefont{Coulais}},
  \emph{\bibinfo{title}{Non-reciprocal robotic metamaterials}},
  \bibinfo{journal}{Nat. Commun.} \textbf{\bibinfo{volume}{10}},
  \bibinfo{pages}{4608} (\bibinfo{year}{2019}).

\bibitem{mcdonaldprx}
\bibinfo{author}{\bibfnamefont{A.}~\bibnamefont{McDonald}},
  \bibinfo{author}{\bibfnamefont{T.}~\bibnamefont{Pereg-Barnea}},
  \bibnamefont{and} \bibinfo{author}{\bibfnamefont{A.~A.} \bibnamefont{Clerk}},
  \emph{\bibinfo{title}{Phase-dependent chiral transport and effective
  non-hermitian dynamics in a bosonic kitaev-majorana chain}},
  \bibinfo{journal}{Phys. Rev. X} \textbf{\bibinfo{volume}{8}},
  \bibinfo{pages}{041031} (\bibinfo{year}{2018}).

\bibitem{kaiwang}
\bibinfo{author}{\bibfnamefont{K.}~\bibnamefont{Wang}},
  \bibinfo{author}{\bibfnamefont{A.}~\bibnamefont{Dutt}},
  \bibinfo{author}{\bibfnamefont{K.~Y.} \bibnamefont{Yang}},
  \bibinfo{author}{\bibfnamefont{C.~C.} \bibnamefont{Wojcik}},
  \bibinfo{author}{\bibfnamefont{J.}~\bibnamefont{Vučković}},
  \bibnamefont{and} \bibinfo{author}{\bibfnamefont{S.}~\bibnamefont{Fan}},
  \emph{\bibinfo{title}{Generating arbitrary topological windings of a
  non-hermitian band}}, \bibinfo{journal}{Science}
  \textbf{\bibinfo{volume}{371}}(\bibinfo{number}{6535}), \bibinfo{pages}{1240}
  (\bibinfo{year}{2021}).

\bibitem{hofmann}
\bibinfo{author}{\bibfnamefont{T.}~\bibnamefont{Hofmann}},
  \bibinfo{author}{\bibfnamefont{T.}~\bibnamefont{Helbig}},
  \bibinfo{author}{\bibfnamefont{F.}~\bibnamefont{Schindler}},
  \bibinfo{author}{\bibfnamefont{N.}~\bibnamefont{Salgo}},
  \bibinfo{author}{\bibfnamefont{M.}~\bibnamefont{Brzezi\ifmmode~\acute{n}\else
  \'{n}\fi{}ska}}, \bibinfo{author}{\bibfnamefont{M.}~\bibnamefont{Greiter}},
  \bibinfo{author}{\bibfnamefont{T.}~\bibnamefont{Kiessling}},
  \bibinfo{author}{\bibfnamefont{D.}~\bibnamefont{Wolf}},
  \bibinfo{author}{\bibfnamefont{A.}~\bibnamefont{Vollhardt}},
  \bibinfo{author}{\bibfnamefont{A.}~\bibnamefont{Kaba\ifmmode~\check{s}\else
  \v{s}\fi{}i}}, \bibinfo{author}{\bibfnamefont{C.~H.} \bibnamefont{Lee}},
  \bibinfo{author}{\bibfnamefont{A.}~\bibnamefont{Bilu\ifmmode \check{s}\else
  \v{s}\fi{}i\ifmmode~\acute{c}\else \'{c}\fi{}}}, \emph{et~al.},
  \emph{\bibinfo{title}{Reciprocal skin effect and its realization in a
  topolectrical circuit}}, \bibinfo{journal}{Phys. Rev. Res.}
  \textbf{\bibinfo{volume}{2}}, \bibinfo{pages}{023265} (\bibinfo{year}{2020}).

\bibitem{liu2022}
\bibinfo{author}{\bibfnamefont{Y.~G.~N.} \bibnamefont{Liu}},
  \bibinfo{author}{\bibfnamefont{Y.}~\bibnamefont{Wei}},
  \bibinfo{author}{\bibfnamefont{O.}~\bibnamefont{Hemmatyar}},
  \bibinfo{author}{\bibfnamefont{G.~G.} \bibnamefont{Pyrialakos}},
  \bibinfo{author}{\bibfnamefont{P.~S.} \bibnamefont{Jung}},
  \bibinfo{author}{\bibfnamefont{D.~N.} \bibnamefont{Christodoulides}},
  \bibnamefont{and}
  \bibinfo{author}{\bibfnamefont{M.}~\bibnamefont{Khajavikhan}},
  \emph{\bibinfo{title}{Complex skin modes in non-hermitian coupled laser
  arrays}}, \bibinfo{journal}{Light: Science \& Applications}
  \textbf{\bibinfo{volume}{11}}, \bibinfo{pages}{336} (\bibinfo{year}{2022}).

\bibitem{wu2023}
\bibinfo{author}{\bibfnamefont{Q.-C.} \bibnamefont{Wu}},
  \bibinfo{author}{\bibfnamefont{J.-L.} \bibnamefont{Zhao}},
  \bibinfo{author}{\bibfnamefont{Y.-L.} \bibnamefont{Fang}},
  \bibinfo{author}{\bibfnamefont{Y.}~\bibnamefont{Zhang}},
  \bibinfo{author}{\bibfnamefont{D.-X.} \bibnamefont{Chen}},
  \bibinfo{author}{\bibfnamefont{C.-P.} \bibnamefont{Yang}}, \bibnamefont{and}
  \bibinfo{author}{\bibfnamefont{F.}~\bibnamefont{Nori}},
  \emph{\bibinfo{title}{Extension of noether’s theorem in pt-symmetry systems
  and its experimental demonstration in an optical setup}},
  \bibinfo{journal}{Sci. China Phys. Mech. Astron.}
  \textbf{\bibinfo{volume}{66}}, \bibinfo{pages}{240312}
  (\bibinfo{year}{2023}).

\bibitem{wangnc}
\bibinfo{author}{\bibfnamefont{K.}~\bibnamefont{Wang}},
  \bibinfo{author}{\bibfnamefont{X.}~\bibnamefont{Qiu}},
  \bibinfo{author}{\bibfnamefont{L.}~\bibnamefont{Xiao}},
  \bibinfo{author}{\bibfnamefont{X.}~\bibnamefont{Zhan}},
  \bibinfo{author}{\bibfnamefont{Z.}~\bibnamefont{Bian}},
  \bibinfo{author}{\bibfnamefont{B.~C.} \bibnamefont{Sanders}},
  \bibinfo{author}{\bibfnamefont{W.}~\bibnamefont{Yi}}, \bibnamefont{and}
  \bibinfo{author}{\bibfnamefont{P.}~\bibnamefont{Xue}},
  \emph{\bibinfo{title}{Observation of emergent momentum–time skyrmions in
  parity–time-symmetric non-unitary quench dynamics}}, \bibinfo{journal}{Nat.
  Commun.} \textbf{\bibinfo{volume}{10}}, \bibinfo{pages}{2293}
  (\bibinfo{year}{2019}).

\bibitem{silva}
\bibinfo{author}{\bibfnamefont{A.}~\bibnamefont{Silva}},
  \emph{\bibinfo{title}{Statistics of the work done on a quantum critical
  system by quenching a control parameter}}, \bibinfo{journal}{Phys. Rev.
  Lett.} \textbf{\bibinfo{volume}{101}}, \bibinfo{pages}{120603}
  (\bibinfo{year}{2008}).

\bibitem{peres}
\bibinfo{author}{\bibfnamefont{A.}~\bibnamefont{Peres}},
  \emph{\bibinfo{title}{Stability of quantum motion in chaotic and regular
  systems}}, \bibinfo{journal}{Phys. Rev. A} \textbf{\bibinfo{volume}{30}},
  \bibinfo{pages}{1610} (\bibinfo{year}{1984}).

\bibitem{physrep}
\bibinfo{author}{\bibfnamefont{T.}~\bibnamefont{Gorina}},
  \bibinfo{author}{\bibfnamefont{T.}~\bibnamefont{Prosen}},
  \bibinfo{author}{\bibfnamefont{T.~H.} \bibnamefont{Seligman}},
  \bibnamefont{and} \bibinfo{author}{\bibfnamefont{M.}~\bibnamefont{Znidaric}},
  \emph{\bibinfo{title}{Dynamics of loschmidt echoes and fidelity decay}},
  \bibinfo{journal}{Phys. Rep.} \textbf{\bibinfo{volume}{435}},
  \bibinfo{pages}{33} (\bibinfo{year}{2006}).

\bibitem{goussev}
\bibinfo{author}{\bibfnamefont{A.}~\bibnamefont{Goussev}},
  \bibinfo{author}{\bibfnamefont{R.~A.} \bibnamefont{Jalabert}},
  \bibinfo{author}{\bibfnamefont{H.~M.} \bibnamefont{Pastawski}},
  \bibnamefont{and} \bibinfo{author}{\bibfnamefont{D.~A.}
  \bibnamefont{Wisniacki}}, \emph{\bibinfo{title}{Loschmidt echo}},
  \bibinfo{journal}{Scholarpedia} \textbf{\bibinfo{volume}{7}},
  \bibinfo{pages}{11687} (\bibinfo{year}{2012}).

\bibitem{herviou}
\bibinfo{author}{\bibfnamefont{L.}~\bibnamefont{Herviou}},
  \bibinfo{author}{\bibfnamefont{N.}~\bibnamefont{Regnault}}, \bibnamefont{and}
  \bibinfo{author}{\bibfnamefont{J.~H.} \bibnamefont{Bardarson}},
  \emph{\bibinfo{title}{{Entanglement spectrum and symmetries in non-Hermitian
  fermionic non-interacting models}}}, \bibinfo{journal}{SciPost Phys.}
  \textbf{\bibinfo{volume}{7}}, \bibinfo{pages}{69} (\bibinfo{year}{2019}).

\bibitem{polkovnikovrmp}
\bibinfo{author}{\bibfnamefont{A.}~\bibnamefont{Polkovnikov}},
  \bibinfo{author}{\bibfnamefont{K.}~\bibnamefont{Sengupta}},
  \bibinfo{author}{\bibfnamefont{A.}~\bibnamefont{Silva}}, \bibnamefont{and}
  \bibinfo{author}{\bibfnamefont{M.}~\bibnamefont{Vengalattore}},
  \emph{\bibinfo{title}{\textit{Colloquium} : Nonequilibrium dynamics of closed
  interacting quantum systems}}, \bibinfo{journal}{Rev. Mod. Phys.}
  \textbf{\bibinfo{volume}{83}}, \bibinfo{pages}{863} (\bibinfo{year}{2011}).

\bibitem{dziarmagareview}
\bibinfo{author}{\bibfnamefont{J.}~\bibnamefont{Dziarmaga}},
  \emph{\bibinfo{title}{Dynamics of a quantum phase transition and relaxation
  to a steady state}}, \bibinfo{journal}{Adv. Phys.}
  \textbf{\bibinfo{volume}{59}}, \bibinfo{pages}{1063} (\bibinfo{year}{2010}).

\bibitem{jing}
\bibinfo{author}{\bibfnamefont{Y.}~\bibnamefont{Jing}},
  \bibinfo{author}{\bibfnamefont{J.-J.} \bibnamefont{Dong}},
  \bibinfo{author}{\bibfnamefont{Y.-Y.} \bibnamefont{Zhang}}, \bibnamefont{and}
  \bibinfo{author}{\bibfnamefont{Z.-X.} \bibnamefont{Hu}},
  \emph{\bibinfo{title}{Biorthogonal dynamical quantum phase transitions in
  non-hermitian systems}}, \bibinfo{note}{{a}rXiv:2307.02993}.

\bibitem{Tang_2022}
\bibinfo{author}{\bibfnamefont{J.-C.} \bibnamefont{Tang}},
  \bibinfo{author}{\bibfnamefont{S.-P.} \bibnamefont{Kou}}, \bibnamefont{and}
  \bibinfo{author}{\bibfnamefont{G.}~\bibnamefont{Sun}},
  \emph{\bibinfo{title}{Dynamical scaling of loschmidt echo in non-hermitian
  systems}}, \bibinfo{journal}{Europhysics Letters}
  \textbf{\bibinfo{volume}{137}}(\bibinfo{number}{4}), \bibinfo{pages}{40001}
  (\bibinfo{year}{2022}).

\bibitem{bender2007}
\bibinfo{author}{\bibfnamefont{C.~M.} \bibnamefont{Bender}},
  \emph{\bibinfo{title}{Making sense of non-hermitian hamiltonians}},
  \bibinfo{journal}{Reports on Progress in Physics}
  \textbf{\bibinfo{volume}{70}}(\bibinfo{number}{6}), \bibinfo{pages}{947}
  (\bibinfo{year}{2007}).

\bibitem{PhysRevB.107.045131}
\bibinfo{author}{\bibfnamefont{A.~N.} \bibnamefont{Poddubny}},
  \emph{\bibinfo{title}{Interaction-induced analog of a non-hermitian skin
  effect in a lattice two-body problem}}, \bibinfo{journal}{Phys. Rev. B}
  \textbf{\bibinfo{volume}{107}}, \bibinfo{pages}{045131}
  (\bibinfo{year}{2023}).

\bibitem{PhysRevB.104.195102}
\bibinfo{author}{\bibfnamefont{C.~H.} \bibnamefont{Lee}},
  \emph{\bibinfo{title}{Many-body topological and skin states without open
  boundaries}}, \bibinfo{journal}{Phys. Rev. B} \textbf{\bibinfo{volume}{104}},
  \bibinfo{pages}{195102} (\bibinfo{year}{2021}).

\bibitem{zhang2022}
\bibinfo{author}{\bibfnamefont{S.-B.} \bibnamefont{Zhang}},
  \bibinfo{author}{\bibfnamefont{M.~M.} \bibnamefont{Denner}},
  \bibinfo{author}{\bibfnamefont{T.~c.~v.}
  \bibnamefont{Bzdu\ifmmode~\check{s}\else \v{s}\fi{}ek}},
  \bibinfo{author}{\bibfnamefont{M.~A.} \bibnamefont{Sentef}},
  \bibnamefont{and} \bibinfo{author}{\bibfnamefont{T.}~\bibnamefont{Neupert}},
  \emph{\bibinfo{title}{Symmetry breaking and spectral structure of the
  interacting hatano-nelson model}}, \bibinfo{journal}{Phys. Rev. B}
  \textbf{\bibinfo{volume}{106}}, \bibinfo{pages}{L121102}
  (\bibinfo{year}{2022}).

\bibitem{alsallom}
\bibinfo{author}{\bibfnamefont{F.}~\bibnamefont{Alsallom}},
  \bibinfo{author}{\bibfnamefont{L.}~\bibnamefont{Herviou}},
  \bibinfo{author}{\bibfnamefont{O.~V.} \bibnamefont{Yazyev}},
  \bibnamefont{and}
  \bibinfo{author}{\bibfnamefont{M.}~\bibnamefont{Brzezi\ifmmode~\acute{n}\else
  \'{n}\fi{}ska}}, \emph{\bibinfo{title}{Fate of the non-hermitian skin effect
  in many-body fermionic systems}}, \bibinfo{journal}{Phys. Rev. Res.}
  \textbf{\bibinfo{volume}{4}}, \bibinfo{pages}{033122} (\bibinfo{year}{2022}).

\bibitem{lee2020}
\bibinfo{author}{\bibfnamefont{E.}~\bibnamefont{Lee}},
  \bibinfo{author}{\bibfnamefont{H.}~\bibnamefont{Lee}}, \bibnamefont{and}
  \bibinfo{author}{\bibfnamefont{B.-J.} \bibnamefont{Yang}},
  \emph{\bibinfo{title}{Many-body approach to non-hermitian physics in
  fermionic systems}}, \bibinfo{journal}{Phys. Rev. B}
  \textbf{\bibinfo{volume}{101}}, \bibinfo{pages}{121109}
  (\bibinfo{year}{2020}).

\bibitem{hamazaki}
\bibinfo{author}{\bibfnamefont{R.}~\bibnamefont{Hamazaki}},
  \bibinfo{author}{\bibfnamefont{K.}~\bibnamefont{Kawabata}}, \bibnamefont{and}
  \bibinfo{author}{\bibfnamefont{M.}~\bibnamefont{Ueda}},
  \emph{\bibinfo{title}{Non-hermitian many-body localization}},
  \bibinfo{journal}{Phys. Rev. Lett.} \textbf{\bibinfo{volume}{123}},
  \bibinfo{pages}{090603} (\bibinfo{year}{2019}).

\bibitem{mu2020}
\bibinfo{author}{\bibfnamefont{S.}~\bibnamefont{Mu}},
  \bibinfo{author}{\bibfnamefont{C.~H.} \bibnamefont{Lee}},
  \bibinfo{author}{\bibfnamefont{L.}~\bibnamefont{Li}}, \bibnamefont{and}
  \bibinfo{author}{\bibfnamefont{J.}~\bibnamefont{Gong}},
  \emph{\bibinfo{title}{Emergent fermi surface in a many-body non-hermitian
  fermionic chain}}, \bibinfo{journal}{Phys. Rev. B}
  \textbf{\bibinfo{volume}{102}}, \bibinfo{pages}{081115}
  (\bibinfo{year}{2020}).

\bibitem{zhangprb2020}
\bibinfo{author}{\bibfnamefont{D.-W.} \bibnamefont{Zhang}},
  \bibinfo{author}{\bibfnamefont{Y.-L.} \bibnamefont{Chen}},
  \bibinfo{author}{\bibfnamefont{G.-Q.} \bibnamefont{Zhang}},
  \bibinfo{author}{\bibfnamefont{L.-J.} \bibnamefont{Lang}},
  \bibinfo{author}{\bibfnamefont{Z.}~\bibnamefont{Li}}, \bibnamefont{and}
  \bibinfo{author}{\bibfnamefont{S.-L.} \bibnamefont{Zhu}},
  \emph{\bibinfo{title}{Skin superfluid, topological mott insulators, and
  asymmetric dynamics in an interacting non-hermitian aubry-andr\'e-harper
  model}}, \bibinfo{journal}{Phys. Rev. B} \textbf{\bibinfo{volume}{101}},
  \bibinfo{pages}{235150} (\bibinfo{year}{2020}).

\bibitem{wang2022}
\bibinfo{author}{\bibfnamefont{Z.}~\bibnamefont{Wang}},
  \bibinfo{author}{\bibfnamefont{L.-J.} \bibnamefont{Lang}}, \bibnamefont{and}
  \bibinfo{author}{\bibfnamefont{L.}~\bibnamefont{He}},
  \emph{\bibinfo{title}{Emergent mott insulators and non-hermitian conservation
  laws in an interacting bosonic chain with noninteger filling and
  nonreciprocal hopping}}, \bibinfo{journal}{Phys. Rev. B}
  \textbf{\bibinfo{volume}{105}}, \bibinfo{pages}{054315}
  (\bibinfo{year}{2022}).

\bibitem{commphyslee}
\bibinfo{author}{\bibfnamefont{R.}~\bibnamefont{Shen}} \bibnamefont{and}
  \bibinfo{author}{\bibfnamefont{C.~H.} \bibnamefont{Lee}},
  \emph{\bibinfo{title}{Non-hermitian skin clusters from strong interactions}},
  \bibinfo{journal}{Communications Physics} \textbf{\bibinfo{volume}{5}},
  \bibinfo{pages}{238} (\bibinfo{year}{2022}).

\bibitem{PhysRevB.106.205147}
\bibinfo{author}{\bibfnamefont{T.}~\bibnamefont{Yoshida}} \bibnamefont{and}
  \bibinfo{author}{\bibfnamefont{Y.}~\bibnamefont{Hatsugai}},
  \emph{\bibinfo{title}{Reduction of one-dimensional non-hermitian point-gap
  topology by interactions}}, \bibinfo{journal}{Phys. Rev. B}
  \textbf{\bibinfo{volume}{106}}, \bibinfo{pages}{205147}
  (\bibinfo{year}{2022}).

\bibitem{giamarchi}
\bibinfo{author}{\bibfnamefont{T.}~\bibnamefont{Giamarchi}},
  \emph{\bibinfo{title}{Quantum Physics in One Dimension}}
  (\bibinfo{publisher}{Oxford University Press}, \bibinfo{address}{Oxford},
  \bibinfo{year}{2004}).

\bibitem{cazalillaboson}
\bibinfo{author}{\bibfnamefont{M.~A.} \bibnamefont{Cazalilla}},
  \emph{\bibinfo{title}{Bosonizing one-dimensional cold atomic gases}},
  \bibinfo{journal}{Journal of Physics B: Atomic, Molecular and Optical
  Physics} \textbf{\bibinfo{volume}{37}}(\bibinfo{number}{7}),
  \bibinfo{pages}{S1} (\bibinfo{year}{2004}).

\bibitem{nersesyan}
\bibinfo{author}{\bibfnamefont{A.~O.} \bibnamefont{Gogolin}},
  \bibinfo{author}{\bibfnamefont{A.~A.} \bibnamefont{Nersesyan}},
  \bibnamefont{and} \bibinfo{author}{\bibfnamefont{A.~M.}
  \bibnamefont{Tsvelik}}, \emph{\bibinfo{title}{Bosonization and Strongly
  Correlated Systems}} (\bibinfo{publisher}{Cambridge University Press},
  \bibinfo{address}{Cambridge}, \bibinfo{year}{1998}).

\bibitem{dorahn}
\bibinfo{author}{\bibfnamefont{B.}~\bibnamefont{D\'ora}} \bibnamefont{and}
  \bibinfo{author}{\bibfnamefont{C.~P.} \bibnamefont{Moca}},
  \emph{\bibinfo{title}{Full counting statistics in the many-body hatano-nelson
  model}}, \bibinfo{journal}{Phys. Rev. B} \textbf{\bibinfo{volume}{106}},
  \bibinfo{pages}{235125} (\bibinfo{year}{2022}).

\bibitem{delft}
\bibinfo{author}{\bibfnamefont{J.}~\bibnamefont{von Delft}} \bibnamefont{and}
  \bibinfo{author}{\bibfnamefont{H.}~\bibnamefont{Schoeller}},
  \emph{\bibinfo{title}{Bosonization for beginners -- refermionization for
  experts}}, \bibinfo{journal}{Ann. Phys. (Leipzig)}
  \textbf{\bibinfo{volume}{7}}, \bibinfo{pages}{225} (\bibinfo{year}{1998}).

\bibitem{glauber}
\bibinfo{author}{\bibfnamefont{R.~J.} \bibnamefont{Glauber}},
  \emph{\bibinfo{title}{Coherent and incoherent states of the radiation
  field}}, \bibinfo{journal}{Phys. Rev.} \textbf{\bibinfo{volume}{131}},
  \bibinfo{pages}{2766} (\bibinfo{year}{1963}).

\bibitem{dmrgmps}
\bibinfo{author}{\bibfnamefont{U.}~\bibnamefont{Schollw{\"o}ck}},
  \emph{\bibinfo{title}{The density-matrix renormalization group in the age of
  matrix product states}}, \bibinfo{journal}{Annals of Physics}
  \textbf{\bibinfo{volume}{326}}, \bibinfo{pages}{96} (\bibinfo{year}{2011}).

\bibitem{itensor}
\bibinfo{author}{\bibfnamefont{M.}~\bibnamefont{Fishman}},
  \bibinfo{author}{\bibfnamefont{S.~R.} \bibnamefont{White}}, \bibnamefont{and}
  \bibinfo{author}{\bibfnamefont{E.~M.} \bibnamefont{Stoudenmire}},
  \emph{\bibinfo{title}{{The ITensor Software Library for Tensor Network
  Calculations}}}, \bibinfo{journal}{SciPost Phys. Codebases}
  p.~\bibinfo{pages}{4} (\bibinfo{year}{2022}).

\bibitem{White-1992}
\bibinfo{author}{\bibfnamefont{S.~R.} \bibnamefont{White}},
  \emph{\bibinfo{title}{Density matrix formulation for quantum renormalization
  groups}}, \bibinfo{journal}{Phys. Rev. Lett.}
  \textbf{\bibinfo{volume}{69}}(\bibinfo{number}{19}), \bibinfo{pages}{2863}
  (\bibinfo{year}{1992}).

\bibitem{gradstein}
\bibinfo{author}{\bibfnamefont{I.}~\bibnamefont{Gradshteyn}} \bibnamefont{and}
  \bibinfo{author}{\bibfnamefont{I.}~\bibnamefont{Ryzhik}},
  \emph{\bibinfo{title}{Table of Integrals, Series, and Products}}
  (\bibinfo{publisher}{Academic Press}, \bibinfo{address}{New York},
  \bibinfo{year}{2007}).

\bibitem{Vidal-2003}
\bibinfo{author}{\bibfnamefont{G.}~\bibnamefont{Vidal}},
  \emph{\bibinfo{title}{Efficient classical simulation of slightly entangled
  quantum computations}}, \bibinfo{journal}{Phys. Rev. Lett.}
  \textbf{\bibinfo{volume}{91}}, \bibinfo{pages}{147902}
  (\bibinfo{year}{2003}).

\bibitem{Note1}
\bibinfo{note}{This can be an artifact of the finite temporal window of Fourier
  transform and the non-perfect periodic behaviour of the generalized Loschmidt
  echo of the lattice model due to high energy modes. This is corroborated by
  the fact that while the $P(W<J)$ is robust with respect to variations of the
  temporal window of Fourier transform, the high energy part in Fig. \ref
  {fig:pw} is very sensitive to even small variations of the time period.}

\end{thebibliography}

\end{document}